\documentclass[aps,prl,preprint,tightenlines,superscriptaddress,showpacs,byrevtex]{revtex4}
\usepackage{graphicx} 
\usepackage{dcolumn}  

\graphicspath{{ps}}

\begin{document}


\preprint{\vbox{ \hbox{BELLE-CONF-0430}
                 \hbox{ICHEP04 8-0677}}}

\title{ \quad\\[0.5cm] 
Measurement of the inclusive charmless semileptonic branching fraction
of $B$ meson using the full reconstruction tag}

\affiliation{Aomori University, Aomori}
\affiliation{Budker Institute of Nuclear Physics, Novosibirsk}
\affiliation{Chiba University, Chiba}
\affiliation{Chonnam National University, Kwangju}
\affiliation{Chuo University, Tokyo}
\affiliation{University of Cincinnati, Cincinnati, Ohio 45221}
\affiliation{University of Frankfurt, Frankfurt}
\affiliation{Gyeongsang National University, Chinju}
\affiliation{University of Hawaii, Honolulu, Hawaii 96822}
\affiliation{High Energy Accelerator Research Organization (KEK), Tsukuba}
\affiliation{Hiroshima Institute of Technology, Hiroshima}
\affiliation{Institute of High Energy Physics, Chinese Academy of Sciences, Beijing}
\affiliation{Institute of High Energy Physics, Vienna}
\affiliation{Institute for Theoretical and Experimental Physics, Moscow}
\affiliation{J. Stefan Institute, Ljubljana}
\affiliation{Kanagawa University, Yokohama}
\affiliation{Korea University, Seoul}
\affiliation{Kyoto University, Kyoto}
\affiliation{Kyungpook National University, Taegu}
\affiliation{Swiss Federal Institute of Technology of Lausanne, EPFL, Lausanne}
\affiliation{University of Ljubljana, Ljubljana}
\affiliation{University of Maribor, Maribor}
\affiliation{University of Melbourne, Victoria}
\affiliation{Nagoya University, Nagoya}
\affiliation{Nara Women's University, Nara}
\affiliation{National Central University, Chung-li}
\affiliation{National Kaohsiung Normal University, Kaohsiung}
\affiliation{National United University, Miao Li}
\affiliation{Department of Physics, National Taiwan University, Taipei}
\affiliation{H. Niewodniczanski Institute of Nuclear Physics, Krakow}
\affiliation{Nihon Dental College, Niigata}
\affiliation{Niigata University, Niigata}
\affiliation{Osaka City University, Osaka}
\affiliation{Osaka University, Osaka}
\affiliation{Panjab University, Chandigarh}
\affiliation{Peking University, Beijing}
\affiliation{Princeton University, Princeton, New Jersey 08545}
\affiliation{RIKEN BNL Research Center, Upton, New York 11973}
\affiliation{Saga University, Saga}
\affiliation{University of Science and Technology of China, Hefei}
\affiliation{Seoul National University, Seoul}
\affiliation{Sungkyunkwan University, Suwon}
\affiliation{University of Sydney, Sydney NSW}
\affiliation{Tata Institute of Fundamental Research, Bombay}
\affiliation{Toho University, Funabashi}
\affiliation{Tohoku Gakuin University, Tagajo}
\affiliation{Tohoku University, Sendai}
\affiliation{Department of Physics, University of Tokyo, Tokyo}
\affiliation{Tokyo Institute of Technology, Tokyo}
\affiliation{Tokyo Metropolitan University, Tokyo}
\affiliation{Tokyo University of Agriculture and Technology, Tokyo}
\affiliation{Toyama National College of Maritime Technology, Toyama}
\affiliation{University of Tsukuba, Tsukuba}
\affiliation{Utkal University, Bhubaneswer}
\affiliation{Virginia Polytechnic Institute and State University, Blacksburg, Virginia 24061}
\affiliation{Yonsei University, Seoul}
  \author{K.~Abe}\affiliation{High Energy Accelerator Research Organization (KEK), Tsukuba} 
  \author{K.~Abe}\affiliation{Tohoku Gakuin University, Tagajo} 
  \author{N.~Abe}\affiliation{Tokyo Institute of Technology, Tokyo} 
  \author{I.~Adachi}\affiliation{High Energy Accelerator Research Organization (KEK), Tsukuba} 
  \author{H.~Aihara}\affiliation{Department of Physics, University of Tokyo, Tokyo} 
  \author{M.~Akatsu}\affiliation{Nagoya University, Nagoya} 
  \author{Y.~Asano}\affiliation{University of Tsukuba, Tsukuba} 
  \author{T.~Aso}\affiliation{Toyama National College of Maritime Technology, Toyama} 
  \author{V.~Aulchenko}\affiliation{Budker Institute of Nuclear Physics, Novosibirsk} 
  \author{T.~Aushev}\affiliation{Institute for Theoretical and Experimental Physics, Moscow} 
% ...
\collaboration{The Belle Collaboration}

\begin{abstract}
We present a preliminary measurement of the inclusive charmless
semileptonic branching fraction of the $B$ meson, based on
140~fb$^{-1}$ of data collected by the Belle detector at the KEKB
$e^+e^-$ asymmetric collider.  Events are tagged by fully
reconstructing one of the $B$ mesons, produced in pairs from
$\Upsilon(4S)$.  The signal for $b \rightarrow u$ semileptonic decay
is distinguished from the $b \rightarrow c$ semileptonic background
using the hadronic and leptonic invariant mass distributions $M_X$ and
$q^2$.  We find the partial branching fraction for the kinematical
region given by $M_X < 1.7~{\rm GeV}/c^2$ and $q^2>8~{\rm GeV}^2/c^2$,
$\Delta{\cal{B}}(B \rightarrow X_u \ell \nu) = [0.99 \pm 0.15 ({\rm
stat})\pm 0.18 ({\rm syst})
\pm 0.04 (b\to u) \pm 0.07 (b\to c)]  \times 10^{-3} $.
Using a theoretical prediction for the extrapolation to the full range of 
$M_X$ and $q^2$ variables, 
we obtain ${\cal{B}}(B \rightarrow X_u \ell \nu)
= [3.37 \pm 0.50 ({\rm stat})\pm 0.60 ({\rm syst}) 
\pm 0.14 (b\to u) \pm 0.24 (b\to c)  \pm 0.50 (f_u~{\rm error})] 
\times 10^{-3}$.
From this measurement, 
the magnitude of the the Cabibbo-Kobayashi-Maskawa matrix element
$V_{ub}$ is $\left|V_{ub}\right|   =  [5.54  \pm
 0.42 ({\rm stat})\pm 0.50 ({\rm syst}) 
\pm 0.12 (b\to u) \pm 0.19 (b\to c) \pm 0.42  (f_u~{\rm error}) 
\pm 0.27 ({\cal{B}} \to |V_{ub}|~{\rm error})]
\times 10^{-3}$.
\end{abstract}

\pacs{12.15.Hh,11.30.Er,13.25.Hw}

\maketitle
\tighten

{\renewcommand{\thefootnote}{\fnsymbol{footnote}}}
\setcounter{footnote}{0}

\section{\boldmath Introduction}
\label{sec:introduction}
Measurement of the Cabibbo-Kobayashi-Maskawa matrix element $|V_{ub}|$
is crucial to test the Standard Model prediction of $CP$ violation.  In this
paper, we report a preliminary measurement of the inclusive charmless
semileptonic branching fraction of the $B$ meson ${\cal{B}}(B
\rightarrow X_u \ell \nu)$, which provides one of the best ways to
determine $|V_{ub}|$. In this measurement, one of the $B$ mesons,
referred to as the tag side meson, $B_{\rm tag}$, is fully
reconstructed in several hadronic decay modes to tag the production,
flavor and charge as well as the momentum of the $B$ meson.  The
semileptonic decay of the other $B$ meson, referred to as the signal
side meson, $B_{\rm sig}$, is detected by the presence of a high
momentum electron or muon.  This method allows one to reconstruct the
invariant mass of the hadronic system in the semileptonic decay,
$M_X$.  The invariant mass squared of the leptonic system, $q^2$, can
also be determined by inferring the missing neutrino momentum.
Both kinematical quantities are used to separate
the $B \rightarrow X_u \ell \nu$ signal decays from the abundant $B
\rightarrow X_c \ell \nu$ background decays, with a good
signal-to-noise ratio.  A similar type of analysis was performed by
BaBar, where only a cut on $M_X$ was applied~\cite{bib:Babar_frec}.
According to a recent theoretical calculation \cite{bib:q2cut}, a
simultaneous cut on the two variables is beneficial because it reduces
theoretical uncertainties.  Such a simultaneous cut on the two
variables was used in a similar analysis by Belle where, instead of the
full reconstruction tagging, a simulated annealing technique was
applied to separate the two $B$ meson decays~\cite{bib:Kakuno}.

This measurement has an advantage over the measurements based on the
lepton momentum alone~\cite{bib:endpoint} in that it covers a
relatively large phase space of the signal spectrum, which reduces the
error in extrapolating the measured partial branching fraction to
obtain the total branching fraction.  While the method has these
unique features, it requires a large sample of $B\overline{B}$ events
because the full reconstruction efficiency is rather low, typically of
the order of 0.3\%.

The data used in the present analysis were collected with the Belle 
detector at the asymmetric energy KEKB~\cite{bib:KEKB} storage ring.
The Belle detector~\cite{bib:BELLE}
 is a large-solid-angle magnetic spectrometer
that consists of a three-layer silicon vertex detector (SVD),
a 50-layer central drift chamber (CDC), 
an array of aerogel threshold \v{C}erenkov counters (ACC),
a barrel-like arrangement of time-of-flight scintillation counters (TOF),
and an electromagnetic calorimeter (ECL) comprised of CsI(Tl) crystals 
located inside a super-conducting solenoid coil 
that provides a 1.5~T magnetic field.  
An iron flux-return located outside of the coil is instrumented
to detect $K_L^0$ mesons and to identify muons (KLM).  

The result presented in this paper is based on a 140~fb${}^{-1}$ data
sample recorded at the $\Upsilon(4S)$ resonance, which contains $152
\times 10^6 B \bar{B}$ pairs.  An additional 15~fb$^{-1}$ data sample
taken at a center-of-mass energy 60\,MeV below the $\Upsilon(4S)$
resonance is used to subtract the background from the $e^+ e^- \to q
\bar{q}$ process ($q=u,\ d,\ s,\ c$).

Monte Carlo simulated events were used to determine the efficiency as
well as signal and background distributions in the control
variables. The simulation program is based on GEANT~\cite{bib:GEANT}
and fully describes the detector geometry and response.  To model the
$B \rightarrow X_u \ell \nu$ decays, we employ a combination of
exclusive channels, where $X_u$ is either a $\pi$ or a
$\rho$~\cite{bib:bu-model-lcsr} or an excited $X_u$
state~\cite{bib:bu-model-isgw2}, and an inclusive model for
non-resonant final states~\cite{bib:bu-model-fn}. The $B \rightarrow
X_c \ell \nu$ transitions are simulated according to the QQ decay
generator~\cite{bib:bc-model}.

This paper is organized as follows. We first discuss the reconstruction
of the tag side $B$ meson, requirements on the signal side and the
signal yield extraction. The observed number of $B \rightarrow X_u
\ell \nu$ decays is then converted into a partial branching fraction
for the kinematical region given by $M_X < 1.7~{\rm GeV}/c^2$ and
$q^2>8~{\rm GeV}^2/c^2$, for which the theoretical uncertainties in
modeling $b \to u$ transitions are small.  The theoretical
result~\cite{bib:q2cut} quoted above together with the $b$-quark shape
function parameters determined from the Belle $b \to s \gamma$
measurement \cite{bib:limosani} is used to extract the branching
fraction for charmless semileptonic decays, which is, finally, used to
obtain the magnitude of the $V_{ub}$ matrix element.

\section{\boldmath Reconstruction of the Tagging Side}
\label{sec:tagging}
In the present analysis,  $B_{\rm tag}$  candidates  
are reconstructed in the decay modes,
$B^0 \to D^{(\ast) -} \pi^+$/$\rho^+$/$a_1^+$/$D^{(\ast)+}_s$ and 
$B^+ \to \bar{D}^{(\ast) 0} \pi^+$/$\rho^+$/$a_1^+$/$D^{(\ast)+}_s$.
Inclusion of charge conjugate decays is implied throughout this paper.

Primary charged tracks are reconstructed with hit information from the
CDC. They are required to satisfy track quality cuts based on their
impact parameters relative to the measured profile of the interaction
point of the two beams.

Charged kaons are identified combining specific ionization ($dE/dx$)
measurements in the CDC, \v{C}erenkov light yields in the ACC and
time-of-flight measurements in the TOF.  The kaon identification
efficiency is approximately $88\%$ and the average pion fake rate is
about $8\%$.

Candidate $\pi^0$ mesons are reconstructed using $\gamma\gamma$ pairs
with an invariant mass between 117.8 and 150.2\,MeV/$c^2$.  Each
photon is required to have a minimum energy deposit of $E_{\gamma}
\geq 50$\,MeV ($E_{\gamma} \geq 30$\,MeV for neutral pions from
$D^{\ast}$ decays).

$K_S^0$ mesons are reconstructed using pairs of charged tracks that
have an invariant mass within $\pm 30$\,MeV/$c^2$ of the known $K_S^0$
mass and a well reconstructed vertex that is displaced from the
interaction point.
Candidate $\rho^+$ and $\rho^0$ mesons are reconstructed in the decay modes
$\pi^+ \pi^0$ and $\pi^+ \pi^-$, by requiring their invariant mass to
be within $\pm 225$\,MeV/$c^2$ of the nominal $\rho$ mass.
Then, $a_1^+$ candidates are selected by combining a $\rho^0$ candidate and
a pion, if their invariant mass lies between 0.7 and 1.6\,GeV/$c^2$
and if the three tracks form a good vertex.

$\bar{D}^0$ candidates are reconstructed in the 
$\bar{D}^0 \to K^+ \pi^-$, $K^+ \pi^- \pi^0$, 
$K^+ \pi^+ \pi^- \pi^-$, $K_S^0 \pi^0$, $K_S^0 \pi^+ \pi^-$, 
$K_S^0 \pi^+ \pi^- \pi^0$ and $K^+ K^-$ decay modes, while 
$D^-$ candidates are reconstructed in the $D^- \to K^+ \pi^- \pi^-$,
$K^+ \pi^- \pi^- \pi^0$, $K_S^0 \pi^-$, $K_S^0 \pi^- \pi^0$,
$K_S^0 \pi^- \pi^- \pi^+$ and $K^+ K^- \pi^-$
decays.
$D_s^+$ candidates are reconstructed in the decay modes 
$D_s^+ \to K_S^0 K^+$ and $K^+ K^- \pi^+$.
These candidates are required to have an invariant mass $m_D$ within
$\pm 4-5\sigma$ of the nominal $D$ mass, where the mass resolution 
$\sigma$ depends on the decay mode. 
$\bar{D}^{\ast}$ mesons are reconstructed by combining the 
$\bar{D}$ candidate and a low momentum pion, 
$D^{\ast -} \to \bar{D}^0 \pi^-$/$D^- \pi^0$ and  
$\bar{D}^{\ast 0} \to \bar{D}^0 \pi^0$.
They are required to have a mass difference
$\Delta m = m_{\bar{D}\pi} - m_{\bar{D}}$ 
within  $\pm 5$~MeV$/c^2$ ($\pm 4-6\sigma$) of the nominal value.    
For the decays with a photon,
 $\bar{D}^{\ast 0} \to \bar{D}^0 \gamma$ and $D_s^{\ast +} 
\to D_s^+ \gamma$, we require that the mass difference
$\Delta m = m_{\bar{D}\gamma} - m_{\bar{D}}$ be  
within $\pm 20$~MeV$/c^2$ ($\pm 2\sigma$) of the nominal value.

The selection of $B_{\rm tag}$ candidates is based on the
beam-constrained mass, $M_{\rm bc} = \sqrt{E_{beam}^{\ast 2}/c^4 -
p^{\ast 2}_{B}/c^2}$, and the energy difference, $\Delta E =
E^{\ast}_{B} - E^{\ast}_{beam}$.  Here $E^{\ast}_{beam}=\sqrt{s}/2
\simeq 5.290$\,GeV is the beam energy in the center of mass system,
and $p^{\ast}_B$ and $E^{\ast}_B$ are the cms momentum and energy of
the reconstructed $B$ meson. (Throughout this paper the variables
calculated in the center of mass system will be denoted with an
asterisk.)  Events satisfying $M_{\rm bc} \geq 5.22$\,GeV/$c^2$ and
$|\Delta E|
\leq 0.3$\,GeV are subject to further analysis.

The combinatorial background from jet-like $e^+e^- \to q\bar{q}$
processes is suppressed by event topology cuts based on the normalized
second Fox-Wolfram moment ($R_2$)~\cite{bib:R2}, $R_2<0.5$, and for
some modes also by a cut on $|\cos \theta_{\rm thrust}|<0.8$, where
$\theta_{\rm thrust}$ is the angle between the thrust axis of the $B$
candidate and that of the rest of the event.

The signal region for the tagging $B$ is defined with the cuts
$M_{\rm bc} \geq 5.26$\,GeV/$c^2$ and $-0.2$\,GeV$< \Delta E < 0.05$\,GeV.
The cuts are optimized to maximize the statistics of the signal, while
minimizing the migration of tracks between the tag and signal sides.

If an event has multiple $B_{\rm tag}$ candidates, we choose the
candidate having the smallest $\chi^2$ based on the deviations from
the nominal values of $\Delta E$, $m_D$, and $\Delta m$ if applicable.

\begin{figure}[t]
\centerline{
\includegraphics[width=0.54\textwidth]{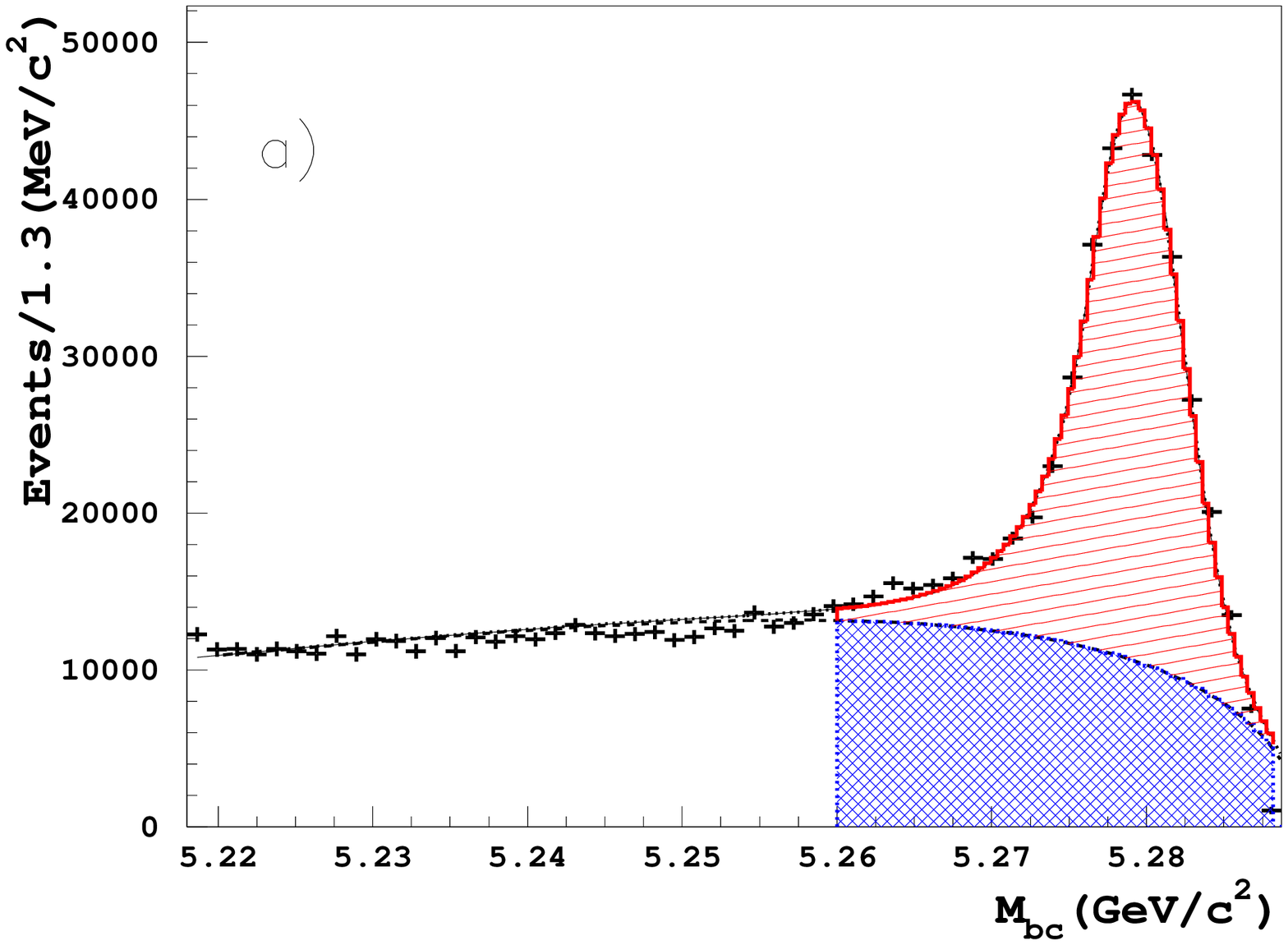}
\includegraphics[width=0.54\textwidth]{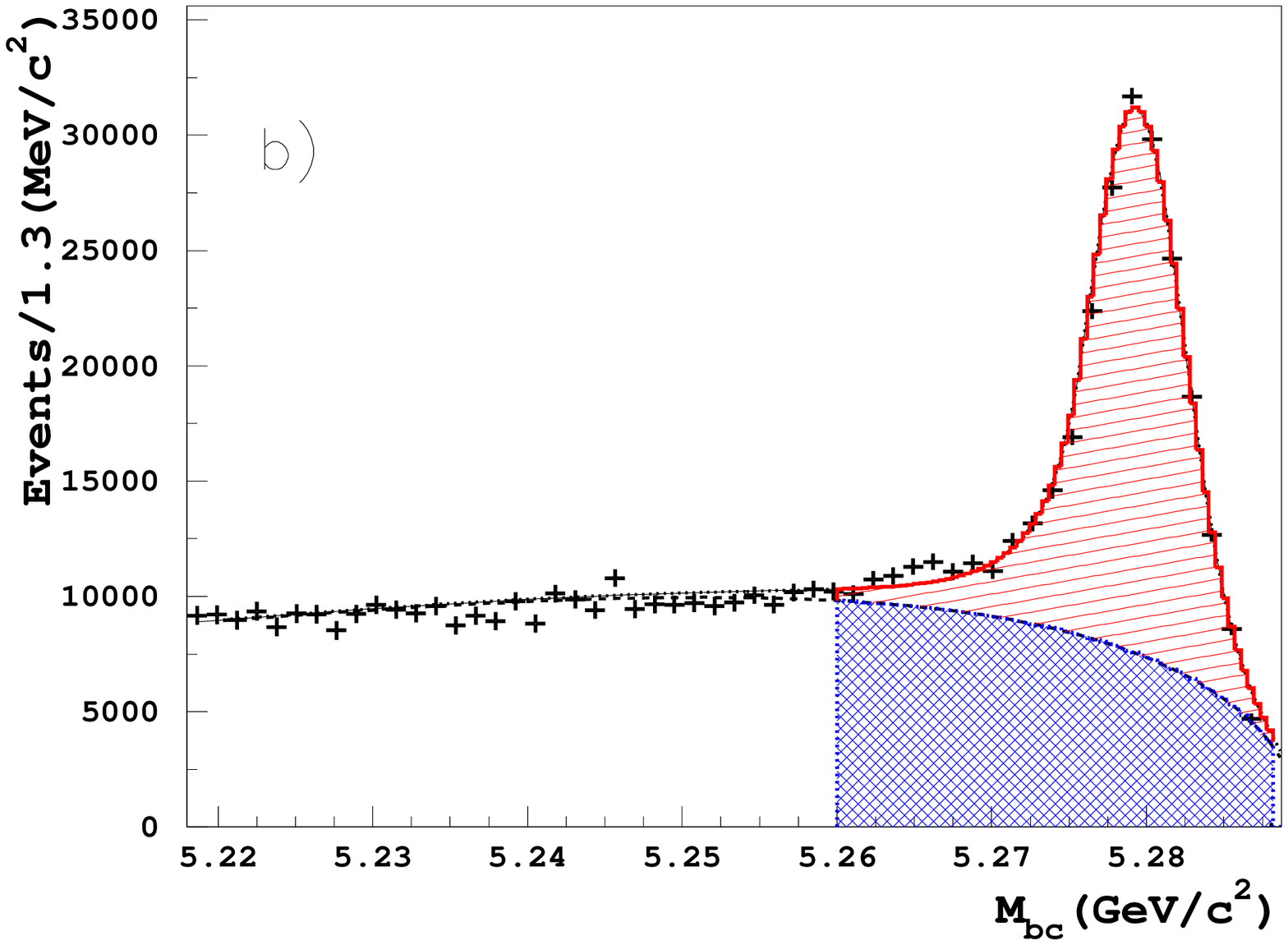}
}
\caption{
Beam-constrained mass ($M_{\rm bc}$) distribution for the $B^+$(a) and
$B^0$(b) candidates with the $-0.2$\,GeV$< \Delta E < 0.05$\,GeV
requirement. The scaled distribution of off-resonance data is
subtracted. The shaded areas indicate the results of the fit in the
$M_{\rm bc}$ signal region.}
\label{fig:Mb}
\end{figure}

Figure~\ref{fig:Mb} shows the distribution of $M_{\rm bc}$ for the
$B^0$ and $B^+$ candidates in the $\Delta E$ signal region.  The
$q\bar{q}$ background contribution is subtracted using the scaled
off-resonance data. The numbers of tagged events are estimated to be
$(1.58 \pm 0.18) \times 10^5$ for $B^0$ and $(2.47 \pm 0.22)
\times 10^5$ for $B^+$, by fitting the distribution with empirical
signal~\cite{bib:cball} and background functions~\cite{bib:argus}. The
reconstruction efficiencies are 0.21\% and 0.33\%, while the purities
are 47\% and 50\% for the $B^0$ and $B^+$ samples, as determined using scaled
off-resonance data subtraction.

\section{\boldmath Reconstruction of the Signal Side}
\label{sec:signalside}
For events tagged by fully reconstructed $B_{\rm tag}$ candidates,
we search for electrons or muons from semileptonic decays of the 
signal side $B$ meson.
Electron identification is based on a combination of the $dE/dx$ value
as measured in the CDC, the response of the ACC, the shower shape in the ECL and 
the ratio of the energy deposit in the ECL to the momentum 
measured by the tracking system.
Muon identification by the KLM relies on the number of hits in
resistive plate counters interspersed in the iron yoke.
The lepton identification efficiencies are about 90\% for both electrons 
and muons in the momentum region above 1\,GeV/$c$.
The hadron fake rate is measured using $K_S^0 \to \pi^+ \pi^-$ and
$\phi^0 \to K^+K^-$ decays, and found to be less than 0.2\% for
electrons and 1.5\% for muons in the same momentum region.

We select electrons having $p^{\ast} \geq 0.6$\,GeV/$c$ and $p_t \geq
0.6$\,GeV/$c$, and muons with $p^{\ast}_{\mu} \geq 0.8$\,GeV/$c$ and
$p_t \geq 0.7$\,GeV/$c$, with $p_t$ being the transverse momentum
component with respect to the positron beam axis.  We also require
that they are detected in the laboratory polar angular region of
$26^{\circ} \leq \theta \leq 140^{\circ}$.  Backgrounds from $J/\psi$
decays, photon conversions in the material of the detector and $\pi^0$
Dalitz decays are minimized by imposing veto conditions; we calculate
invariant masses for each lepton candidate when combined with opposite
charge leptons ($m_{\ell\ell}$) and with an additional photon in the
case of electrons ($m_{ee\gamma}$), and reject the lepton if
$m_{\ell\ell}$ lies within $\pm 5\sigma$ of the nominal $J/\psi$ mass,
$m_{ee}$ within $\pm100$\,MeV/$c^2$ or $m_{ee\gamma}$ within $\pm
3\sigma$ of the nominal $\pi^0$ mass.

For the prompt semileptonic decay signal, we require a lepton with
momentum $p^{\ast}$ exceeding 1\,GeV/$c$.  We also require the lepton
charge to be consistent with a prompt semileptonic decay, when the
$B_{\rm tag}$ candidate is charged.  No requirement is imposed
on the lepton charge when the $B_{\rm tag}$ candidate is neutral.

\begin{figure}[t]
\centerline{
\includegraphics[width=0.54\textwidth]{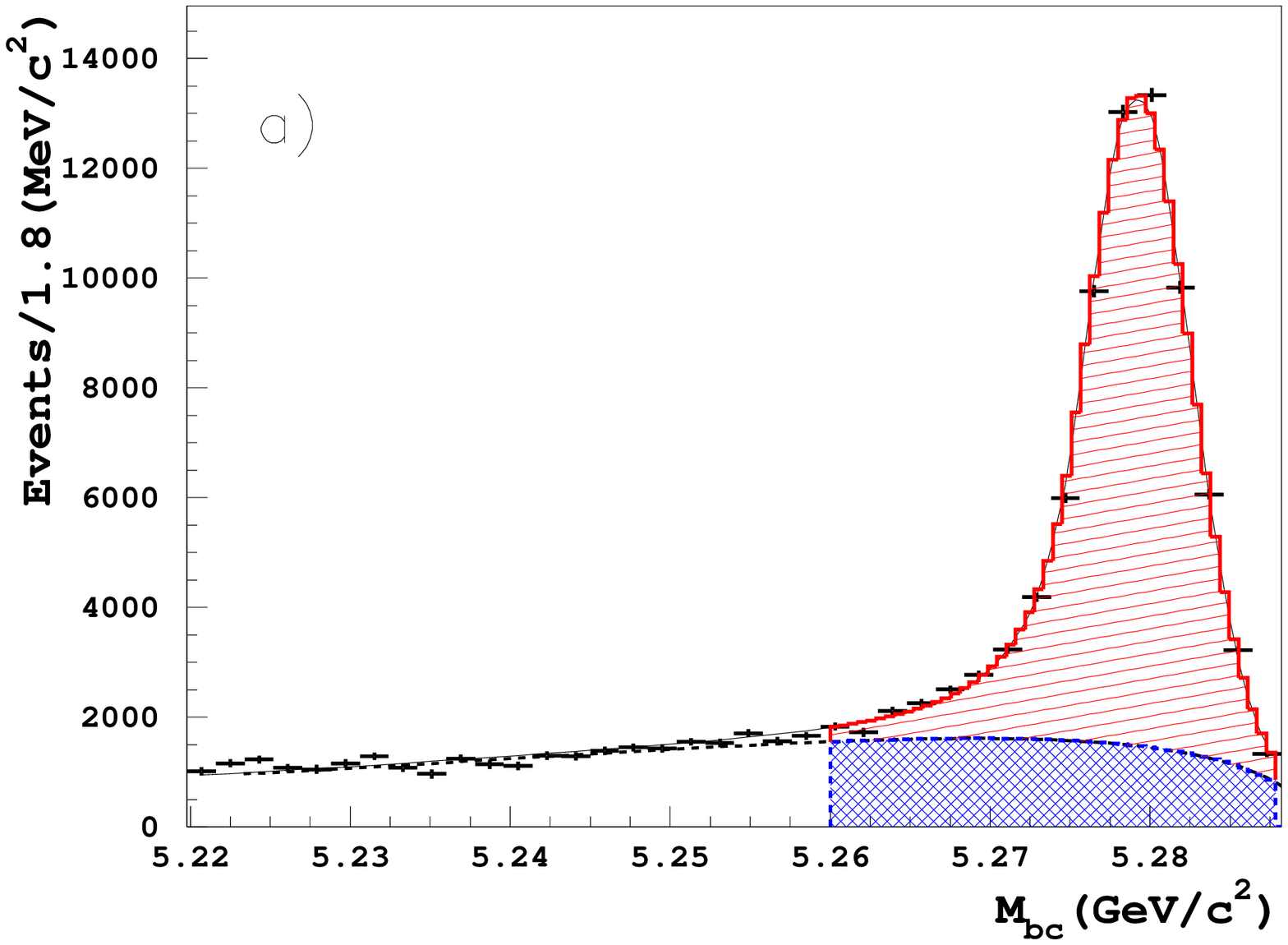}
\includegraphics[width=0.54\textwidth]{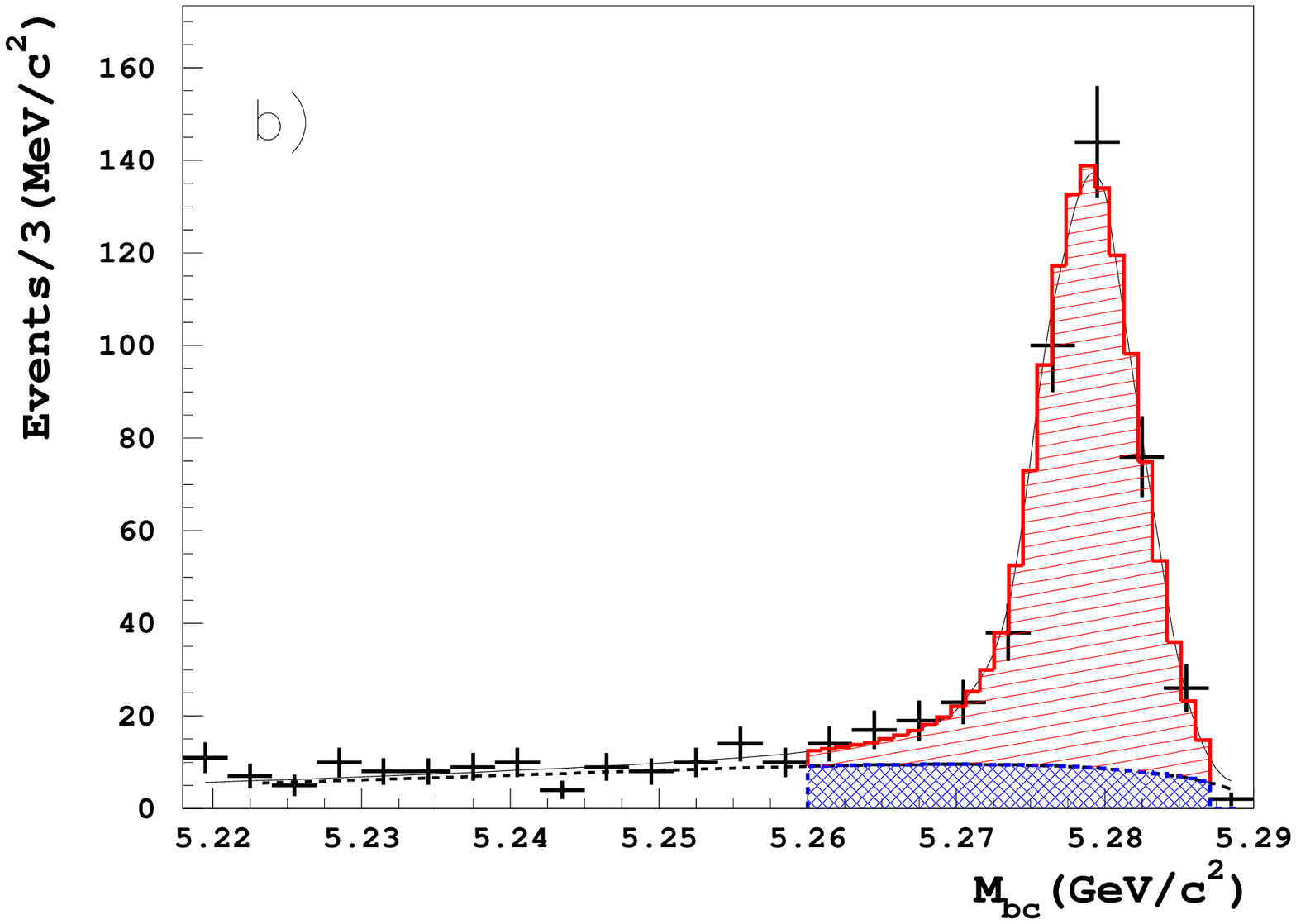}
}
\caption{Beam-constrained mass ($M_{\rm bc}$) distribution 
for prompt semileptonic decays (a) and events satisfying all $B
\rightarrow X_u \ell \nu$ signal requirements, including $M_X <
1.7~{\rm GeV}/c^2$ and $q^2>8~{\rm GeV}^2/c^2$ (b). The shaded areas
indicate the results of the fit in the $M_{\rm bc}$ signal region.  }
\label{fig:Mb-sig}
\end{figure}
 
The $B \rightarrow X_u \ell \nu$ signal events are selected by
imposing several requirements to suppress the $B \rightarrow X_c \ell
\nu$ background as well as the background from poorly reconstructed
events.  Fake charged tracks, arising mainly from duplication in the
tracking of low momentum curling tracks, are removed on the basis of
the angle between the two track candidates, and the difference in
their momenta.
We require that the event contains only one lepton, has zero net
charge ($\sum_i Q_i = 0$) and has a missing mass consistent with zero
($-1$\,GeV$^2/c^4 \leq m_{miss}^2 \leq 0.5$\,GeV$^2$/$c^4$). In order
to further suppress the $B \rightarrow X_c \ell \nu$ background, we
require that the number of kaons, either $K^\pm$ or $K_S^0$, is zero
($N_K = 0$) on the signal side.  To reject events containing a $K_L$
meson, we require that the angle between the missing momentum and the
direction of the candidate $K_L$ cluster be greater than $37^\circ$.

Finally, we select the signal events with requirements on the
invariant mass of the hadronic system $M_X$ and invariant mass
squared of the leptonic system $q^2$. The variable $M_X$ is calculated
from the measured momenta of all charged tracks and energy deposits of
all neutral clusters in the ECL that are not used in the $B_{\rm tag}$
reconstruction or not identified as leptons.  Charged tracks are
assigned the mass of the pion, kaon or proton, based on the
information from the particle identification system.
According to Monte Carlo simulations the resolution in $M_X$ for the
selected events is found to be about 125\,MeV/$c^2$ and 130\,MeV/$c^2$
for $B \to X_u \ell \nu$ and $B \to X_c \ell \nu$ processes, respectively.
The four-momentum transfer $q$ is calculated as $q = p_{\Upsilon(4S)}
- p_{B_{\rm tag}} - p_{X}$, where $p_{\Upsilon(4S)}$, $p_{B_{\rm
tag}}$ and $p_{X}$ are the four-momentum vectors of the
$\Upsilon(4S)$, $B_{\rm tag}$ and the reconstructed hadronic system,
respectively.

Our signal region is defined as $M_X < 1.7~{\rm GeV}/c^2$ and
$q^2>8~{\rm GeV}^2/c^2$. With these requirements, we suppress the
kinematic region where the theoretical interpretation of the result
would be difficult~\cite{bib:q2cut}, and also reduce the level of the
$B \rightarrow X_c \ell \nu$ background.  The distribution in the beam
constrained mass for the selected events is shown in
Fig.~\ref{fig:Mb-sig}(b).

\section{\boldmath Signal Yield Extraction}
\label{sec:signalyield}

\begin{figure}[t]
\centerline{
\includegraphics[width=0.54\textwidth]{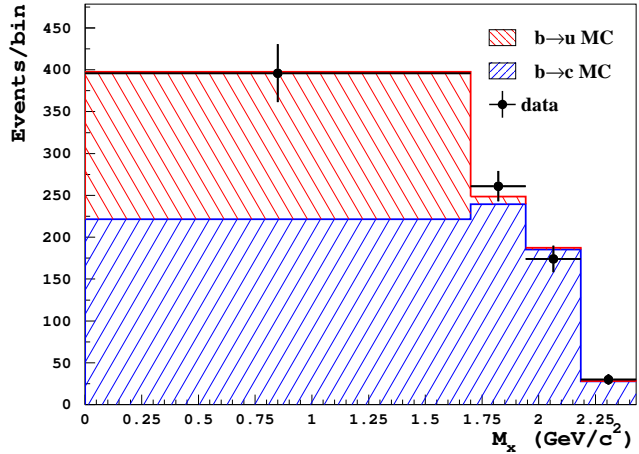}}
\caption{
$M_X$ distribution for the selected events with $q^2>8$~GeV$^2/c^2$,
data and fitted contributions of the $B \rightarrow X_c \ell \nu$ and
$B \rightarrow X_u \ell \nu$ transitions.}
\label{fig:MX-fewbin}
\end{figure}

The signal yield is extracted by a fit to the distribution of hadronic
invariant mass $M_X$ for the selected events with $q^2>8~{\rm
GeV}^2/c^2$. The distribution of events in hadronic mass is
determined by dividing the data into several $M_{X}$ bins. For each
bin, the yield is extracted from a fit with empirical
signal~\cite{bib:cball} and background functions~\cite{bib:argus} to
the corresponding $M_{\rm bc}$ distribution, an example of which is
shown in Fig.~\ref{fig:Mb-sig}(b). Figure~\ref{fig:MX-fewbin} shows
the resulting $M_X$ distribution.

\begin{figure}[t]
\centerline{
\includegraphics[width=0.54\textwidth]{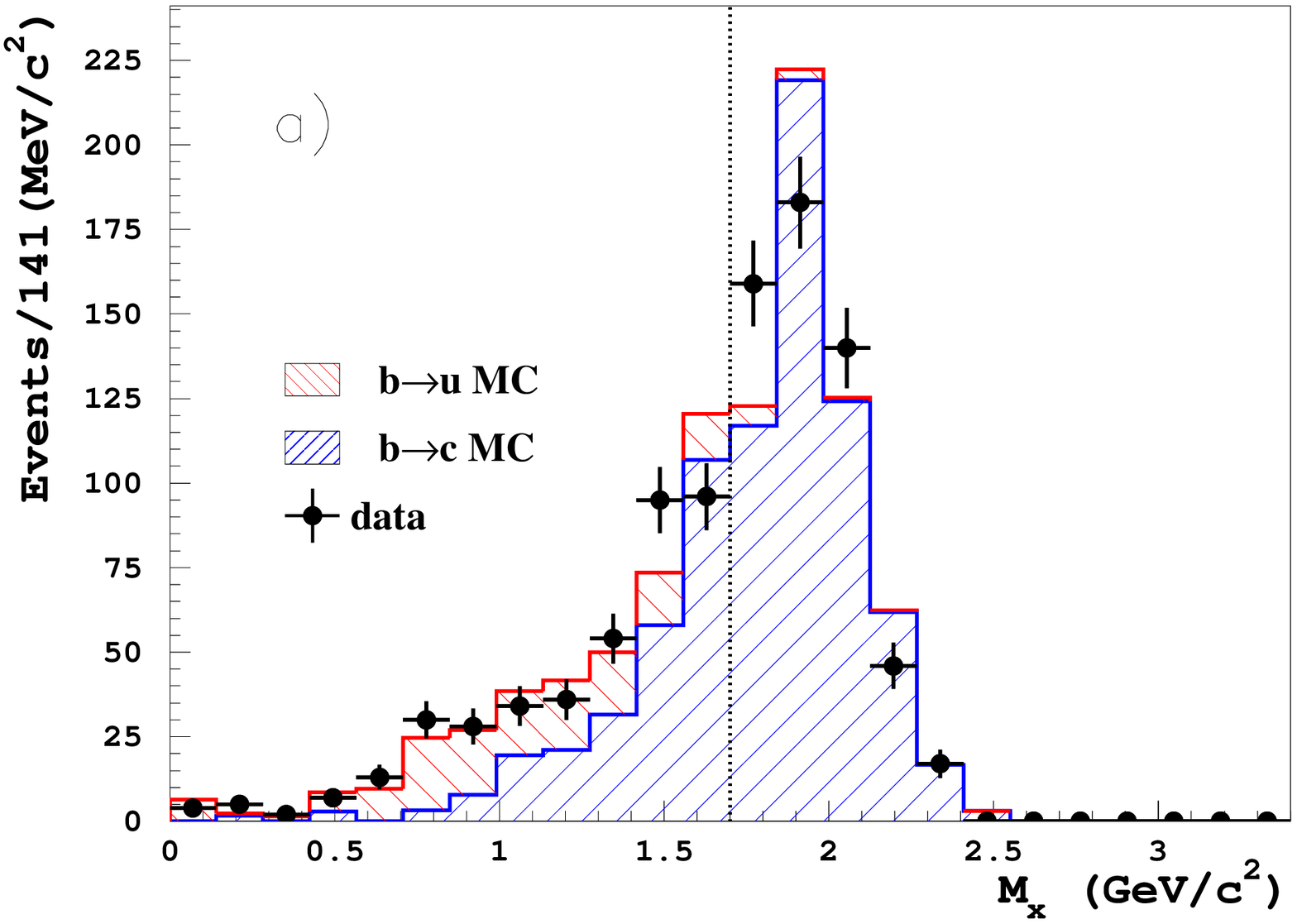}
\includegraphics[width=0.54\textwidth]{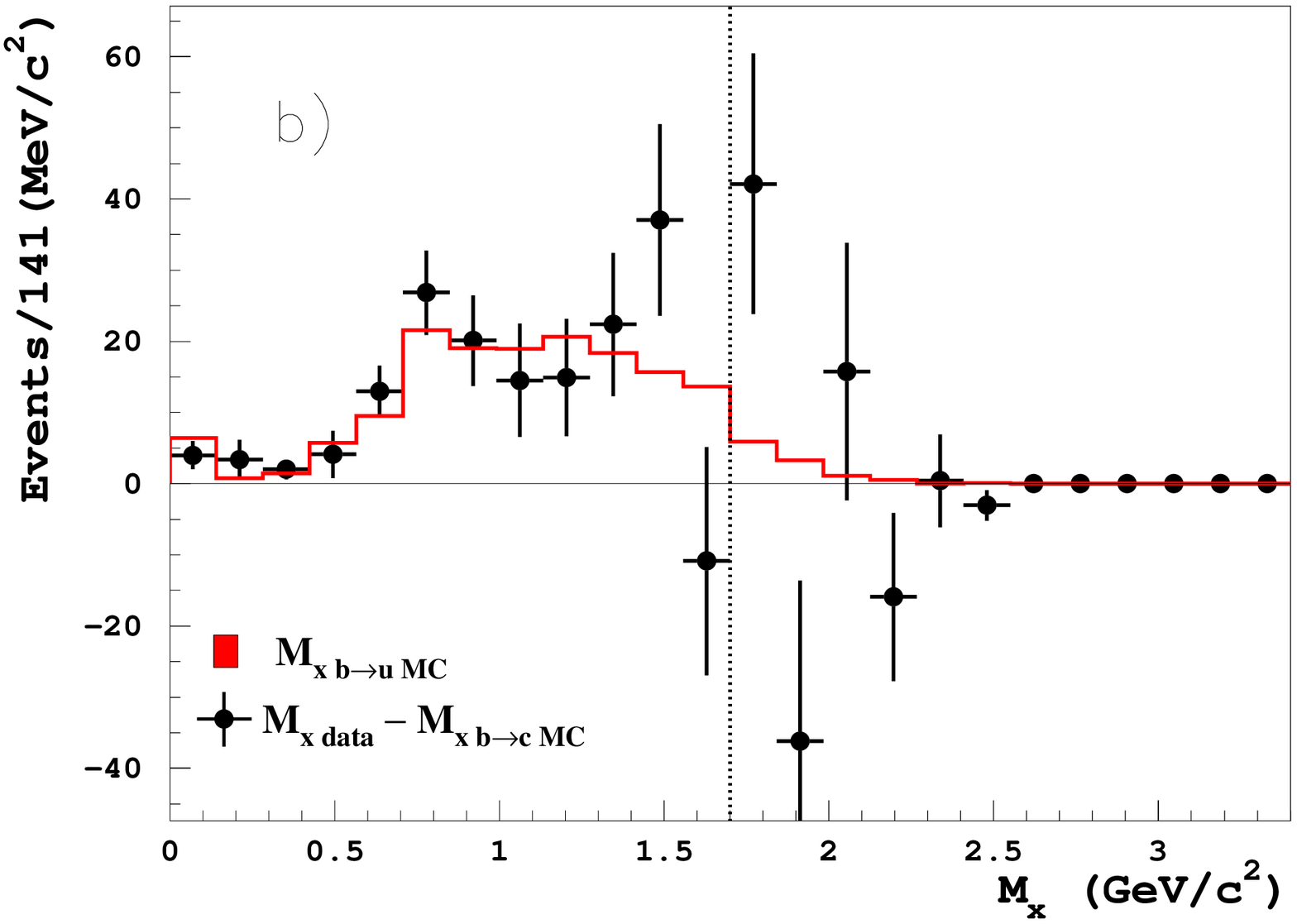}
}
\caption{
$M_X$ distributions for the selected events with $q^2>8$~GeV$^2/c^2$:
a) data and fitted contributions of the $B \rightarrow X_c \ell \nu$
and $B \rightarrow X_u \ell \nu$ transitions, b) measured distribution
for the $B \rightarrow X_u \ell \nu$ transition, obtained by
subtracting the fitted $B \rightarrow X_c \ell \nu$ contribution from
the data, compared to the simulated $B \rightarrow X_u \ell \nu$
contribution with the normalization given by the fit. The final cut on
$M_X$ is indicated by a dotted line.}
\label{fig:MX}
\end{figure}

\begin{figure}[t]
\centerline{
\includegraphics[width=0.54\textwidth]{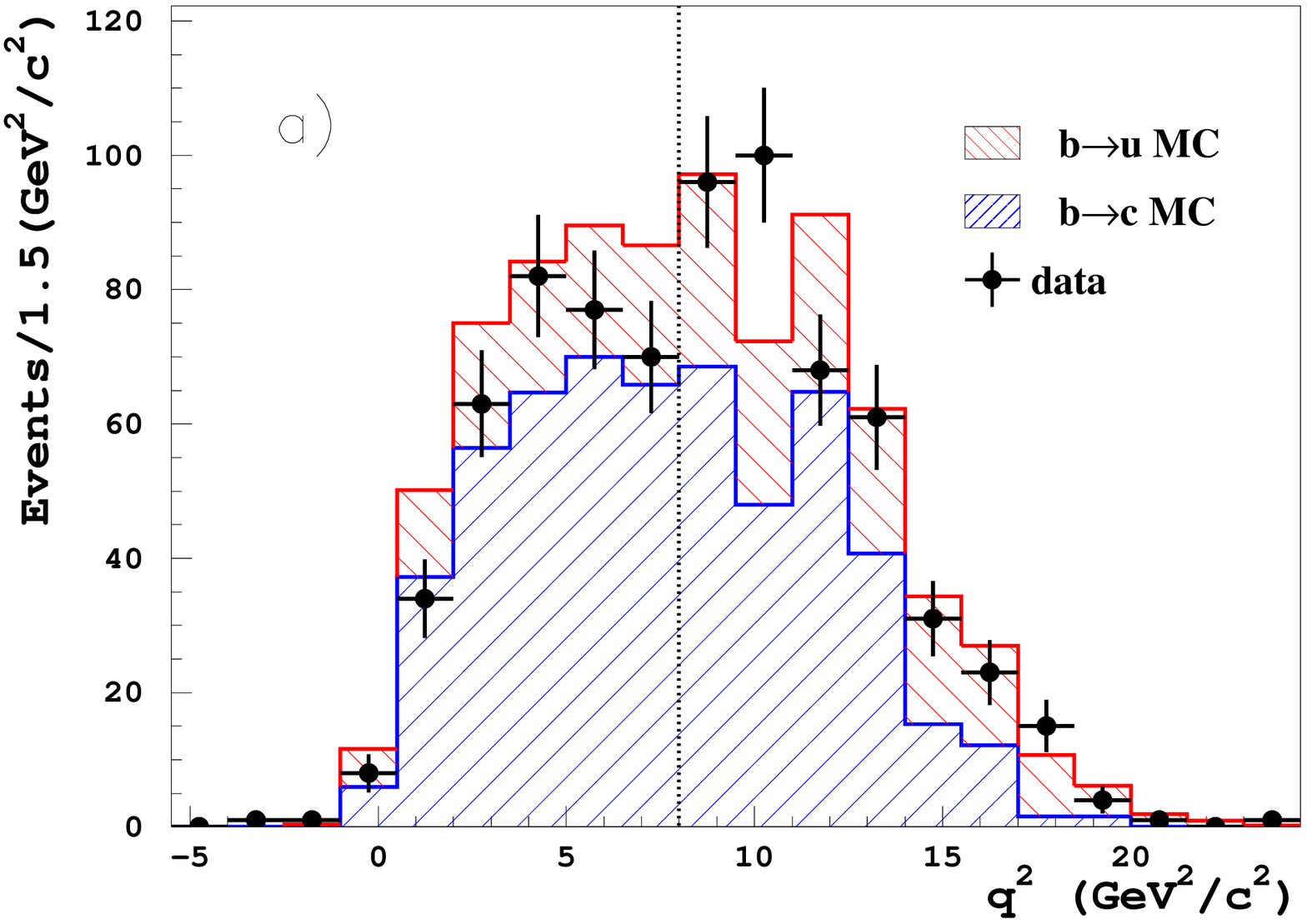}
\includegraphics[width=0.54\textwidth]{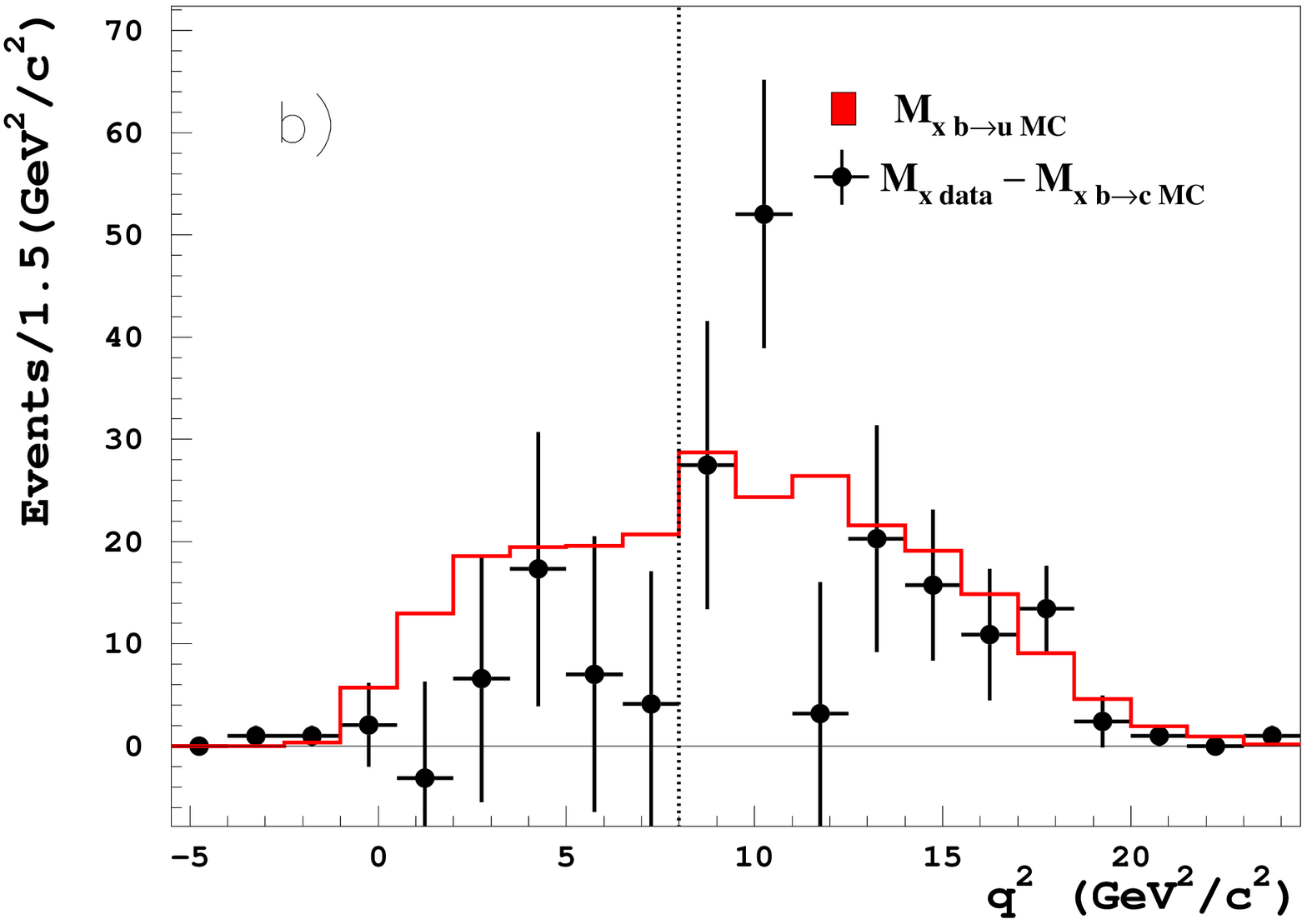}}
\caption{
The $q^2$ distribution for the selected events with
$M_X<1.7$~GeV/$c^2$: a) data and contributions of the $B \rightarrow
X_c \ell \nu$ and $B \rightarrow X_u \ell \nu$ transitions, normalized
with the fit to the $M_X$ distribution, b) measured distribution for
the $B \rightarrow X_u \ell \nu$ transition, obtained by subtracting
the $B \rightarrow X_c \ell \nu$ contribution from the data, compared
to the simulated $B \rightarrow X_u \ell \nu$ contribution. The final
cut on $q^2$ is indicated by a dotted line.}
\label{fig:q2}
\end{figure}

To determine the raw number of events corresponding to the $B
\rightarrow X_u \ell \nu$ process, we first fit the $M_X$ distribution
with two contributions, the signal and the background from the $B
\rightarrow X_c \ell \nu$ process.  The shapes of both contributions
are determined from Monte Carlo simulation.  The result of the fit is
indicated in Fig.~\ref{fig:MX-fewbin}.  Using the same relative
normalization of the processes, we plot in Figs.~\ref{fig:MX} and
\ref{fig:q2} the distributions of events over the $M_X$ (finer
granularity as compared to Fig.~\ref{fig:MX-fewbin}) and $q^2$
variables. In the regions of low $M_X$ ($< 1.7$\,GeV/$c^2$) and high
$q^2$ a clear contribution from $B \rightarrow X_u \ell \nu$ can be
observed.

The raw yield of $B \rightarrow X_u \ell \nu$ decay in the signal
region, $N_{b \rightarrow u}^{\rm raw}=174\pm 26$, is determined by
subtracting the fitted $B \rightarrow X_c \ell \nu$ contribution
(Fig.~\ref{fig:MX-fewbin}). The error is statistical only.

In order to extract the partial branching fraction $\Delta{\cal{B}}$
for $B \rightarrow X_u \ell \nu$ in the signal region $M_X < 1.7~{\rm
GeV}/c^2,q^2>8~{\rm GeV}^2/c^2$, Monte Carlo simulation is used to
convert the observed number of events $N_{b \rightarrow u}^{\rm raw}$
to the true number of signal events produced in this region, $N_{b
\rightarrow u}$, and to estimate the efficiency for these events to be
observed anywhere.  $N_{b \rightarrow u}$ is estimated by $N_{b
\rightarrow u} = N_{b \rightarrow u}^{\rm raw} \times F$, with $F = 1
+ N_2/N_1 - N_3/N_1$. Here $N_1$ is the number of simulated events
observed in the signal region and $N_2$ ($N_3$) is the number of
events generated inside (outside) the signal region and observed
outside (inside) the signal region. We find $F=0.984 \pm 0.014$, and
thus $N_{b \rightarrow u} = 171 \pm 26$.  The efficiency
$\epsilon_{\rm sel}^{b \rightarrow u}$ for selecting $B \rightarrow
X_u \ell \nu$ events after the lepton momentum cut is predicted to be
$27.4$\%.

The relative partial branching fraction 
$\Delta{\cal{B}}(B \rightarrow X_u \ell \nu)/
{\cal{B}}(B \rightarrow X \ell \nu)$ is obtained by
\begin{equation}
\frac{\Delta{\cal{B}}(B \rightarrow X_u \ell \nu)}
{{\cal{B}}(B \rightarrow X \ell \nu)}
= \frac{N_{b \rightarrow u}}{N_{\rm sl}} \times
\frac{1}{\epsilon_{\rm sel}^{b \rightarrow u}} \times
\frac{\epsilon_{\rm frec}^{\rm sl}}{\epsilon_{\rm frec}^{b \rightarrow u}} \times
\frac{\epsilon_{l}^{\rm sl}}{\epsilon_{l}^{b \rightarrow u}}~~~~.
\end{equation}
Here $N_{\rm sl}=(5.07 \pm 0.04) \times 10^4$ is the number of events
having at least one lepton with $p^{\ast} \geq 1.0$\,GeV/$c$,
determined from a fit to the corresponding $M_{\rm bc}$ distribution
(Fig.~\ref{fig:Mb-sig}(a)), and corrected for the expected fraction of
background events from non-semileptonic decays (14.1\%),as estimated
by MC simulation.
The factor $\epsilon_{\rm frec}^{sl}/\epsilon_{\rm frec}^{b
\rightarrow u}$ accounts for a possible difference in the $B_{\rm
tag}$ reconstruction efficiency in the presence of a semileptonic or
$B \rightarrow X_u \ell \nu$ decay; $\epsilon_{l}^{\rm
sl}/\epsilon_{l}^{b \rightarrow u}$ is the ratio of fractions of
semileptonic decay leptons with $p^{\ast} > 1$\,GeV/$c$, in the whole
phase space for the $B \rightarrow X_c \ell \nu$ events, and within
the region $M_X < 1.7~{\rm GeV}/c^2$ and $q^2>8~{\rm GeV}^2/c^2$ for signal
events.
The product of efficiency ratios is found to be $\epsilon_{\rm frec}^{\rm
sl}/\epsilon_{\rm frec}^{b \rightarrow u} \times \epsilon_{l}^{\rm
sl}/\epsilon_{l}^{b \rightarrow u}=0.75 \pm 0.048$.

\section{\boldmath Systematic Errors}
\label{sec:systematic}

The major sources of systematic error in the relative branching
fraction are the uncertainty in the background subtraction to extract
the yield $N_{b \rightarrow u}$, the uncertainties in the calculation
of the efficiency ratios, and the uncertainty due to the treatment of
$B \rightarrow X_u \ell \nu$ decays in the Monte Carlo simulation that
is used to estimate $\epsilon_{\rm sel}^{b \rightarrow u}$ and $F$.

The validity of the $B \rightarrow X_c \ell \nu$ background simulation
is tested with the $B \rightarrow X_c \ell \nu$ enhanced control
sample, where all selection cuts are applied with one exception: we
require at least one kaon in the event.  We have checked that for the
$M_X$ distribution of this control sample there is good agreement
between data and the simulation.  The relative errors arising from
uncertainties in the background subtraction to extract $N_{b
\rightarrow u}$ are estimated to be $7$\% due to the uncertainty in
the modeling of the $B \rightarrow X_c \ell \nu$ background and $15$\%
due to the limited statistics for the simulated background events.

Statistical uncertainty in the determination of efficiency ratios
contributes 6\% to the systematic error.  The uncertainty in the ratio
$F / \epsilon_{\rm sel}^{b \rightarrow u}$ is estimated by varying the
parameters of the simulation model for $B \rightarrow X_u \ell
\nu$ decays.  We assign a 4\% error after varying the parameters
of the inclusive model within their errors.

The uncertainties  due to systematic errors in tracking efficiency,
particle identification efficiency and cluster finding efficiency
are estimated by varying them by their respective errors, and 
observing the effect on the value of partial branching fraction.
These sources give correlated errors on simulated $b \rightarrow u$
and $b \rightarrow c$ events so we add or subtract the error of each
source linearly for the two samples, according to the relative sign of
their effect. The contributions for different sources are combined in
quadrature to a systematic error of $6.5$\%. The total systematic
error excluding $b \to u$ and $b \to c$ model dependences amounts to 18\%.

\section{\boldmath Summary}
\label{sec:summary}

In summary, the relative partial branching fraction 
for $B \rightarrow X_u \ell \nu$ decays in the kinematic region 
$M_X < 1.7~{\rm GeV}/c^2,q^2>8~{\rm GeV}^2/c^2$ is 
\begin{equation}
\frac{\Delta{\cal{B}}(B \rightarrow X_u \ell \nu)}
{{\cal{B}}(B \rightarrow X \ell \nu)} 
= [0.92  \pm 0.14 ({\rm stat})\pm 0.17 ({\rm syst}) 
\pm 0.04 (b\to u) \pm 0.06 (b\to c)]  \times 10^{-2}~~~~,
\end{equation}
where the first error is statistical, the second experimental
systematic, and the third and fourth arise from uncertainties in
modeling the $B \rightarrow X_u \ell \nu$ and $B \rightarrow X_c \ell
\nu$ transitions. By using the measured semileptonic branching
fraction ${\cal{B}}(B \rightarrow X \ell \nu) = 0.1073 \pm
0.0028$~\cite{bib:PDG2004}, we obtain
\begin{equation}
{\Delta\cal{B}}(B \rightarrow X_u \ell \nu) 
= [0.99  \pm 0.15 ({\rm stat})\pm 0.18 ({\rm syst}) 
\pm 0.04 (b\to u) \pm 0.07 (b\to c)]  \times 10^{-3}~~~.
\end{equation}

The branching fraction ${\cal{B}}(B \rightarrow X_u \ell \nu)$ 
is calculated from the above value through the expression
\begin{equation}
{\cal{B}}(B \rightarrow X_u \ell \nu) = 
\Delta{\cal{B}}(B \rightarrow X_u \ell \nu)/f_u~~~~.
\end{equation}
The extrapolation coefficient $f_u$ is estimated to be $0.303\pm
0.035$ using the De Fazio and Neubert (FN) prescription
\cite{bib:bu-model-fn} with the $b$-quark shape function parameters 
$m_b=4.62$~GeV$/c^2$ and $\mu_\pi^2=0.40$~GeV$^2/c^2$ and their one
sigma error ellipse, which are determined from a recent Belle $B \to
X_s \gamma$ measurement \cite{bib:limosani,bib:patrick}. It is
rescaled to $0.294 \pm 0.035$ by multiplying by a factor $f_{u0}({\rm
BLL})/f_{u0}({\rm FN}) =0.324/0.334$, where $f_{u0}({\rm BLL})$ and
$f_{u0}({\rm FN})$ denote the $f_u$ values calculated for the values
of $m_b=4.71$~GeV$/c^2$ and $\mu_\pi^2=0.2$~GeV$^2/c^2$ by the Bauer,
Ligeti and Luke (BLL)~\cite{bib:q2cut} and De Fazio and Neubert
prescriptions, respectively. In both $\alpha_s$ and $1/m_b$ expansions
the BLL prescription contains corrections of higher order than the FN
prescription.  The uncertainty of $f_u$ is modified by including the
contributions from the subleading shape function and the weak
annihilation.  The former is estimated to be
4\%~\cite{bib:hfag-writeup} and the latter is estimated to be
8\%~\cite{bib:q2cut}. As a result we use $f_u=0.294\pm 0.044$ in
Eq.(4). This yields
\begin{eqnarray}
{\cal{B}}(B \rightarrow X_u \ell \nu) = 
[3.37 \pm
0.50 ({\rm stat})\pm
0.60 ({\rm syst})\pm
0.14 (b\to u) \\ \nonumber\pm
0.24 (b\to c) \pm
0.50 (f_u~{\rm error})]
 \times 10^{-3}~~~~,
\end{eqnarray}
where the additional error comes from the uncertainty 
in the calculation of $f_u$.

Combining this result with the average $B$ lifetime 
$\tau_B=(1.587\pm 0.011)$~ps \cite{bib:PDG2004},

the CKM matrix element $|V_{ub}|$ is obtained by using the formula
\begin{equation}
|V_{ub}| = 0.00424~~\left( \frac{{\cal{B}}(B \rightarrow X_u \ell \nu)}{0.002}
\frac{1.61 {\rm ps}}{\tau_B}\right)^{1/2}~~~,
\end{equation}
which is an updated version of the expression given in \cite{bib:PDG2002},
and incorporates the latest measurements
of the heavy-quark parameters~\cite{bib:BaBar_vcb}. 
Finally, we obtain
\begin{eqnarray}
\left|V_{ub}\right|   = [5.54  \pm
 0.42 ({\rm stat})\pm 0.50 ({\rm syst})  
\pm 0.12 (b\to u) \pm 0.19 (b\to c) \\ \nonumber
\pm 0.42  (f_u~{\rm error}) 
\pm 0.27 ({\cal{B}} \to |V_{ub}|~{\rm error})] 
\times 10^{-3}~~~~.
\end{eqnarray}
The first four errors are statistical, systematic, $b\to c$ and $b\to u$ model 
dependence. The latter two are due to uncertainties of $f_u$ and of the relation 
between $\cal{B}$ and $|V_{ub}|$ in Eq.~(6), respectively.

The present work demonstrates the effectiveness of $M_X$ and $q^2$
measurements of $B \rightarrow X_u \ell \nu$ decays
using full reconstruction tagging.
The result is consistent with previous measurements
of ${\cal{B}}(B \rightarrow X_u \ell \nu)$
~\cite{bib:endpoint,bib:inclusive,bib:Babar_frec,bib:Kakuno}.
In the future, further accumulation of data as well as MC data will
allow us to reduce the statistical error, and better understanding of
the signal and the background components will help to improve the
experimental systematics as well as constrain the theoretical
uncertainties.

We thank the KEKB group for the excellent operation of the
accelerator, the KEK Cryogenics group for the efficient
operation of the solenoid, and the KEK computer group and
the National Institute of Informatics for valuable computing
and Super-SINET network support. We acknowledge support from
the Ministry of Education, Culture, Sports, Science, and
Technology of Japan and the Japan Society for the Promotion
of Science; the Australian Research Council and the
Australian Department of Education, Science and Training;
the National Science Foundation of China under contract
No.~10175071; the Department of Science and Technology of
India; the BK21 program of the Ministry of Education of
Korea and the CHEP SRC program of the Korea Science and
Engineering Foundation; the Polish State Committee for
Scientific Research under contract No.~2P03B 01324; the
Ministry of Science and Technology of the Russian
Federation; the Ministry of Education, Science and Sport of
the Republic of Slovenia; the National Science Council and
the Ministry of Education of Taiwan; and the U.S.\
Department of Energy.


\begin{thebibliography}{99}

\bibitem{bib:Babar_frec}
B. Aubert {\it et al.} (BaBar Collaboration), Phys. Rev. Lett. {\bf 92}, 071802 (2004).

\bibitem{bib:q2cut} C. W. Bauer, Z. Ligeti and M. E. Luke, Phys. Rev. {\bf D 64}, 113004  (2001).

\bibitem{bib:Kakuno}
H.~Kakuno {\it et al.} (Belle Collaboration), Phys. Rev. Lett. {\bf 92}, 101801 (2004).

\bibitem{bib:endpoint}
R. Fulton {\it et al.} (CLEO Collaboration), Phys. Rev. Lett. {\bf 64} 16 (1990),
J. Bartelt {\it et al.} (CLEO Collaboration), Phys. Rev. Lett. {\bf 71}, 4111 (1993),
A. Bornheim {\it et al.} (CLEO Collaboration), Phys. Rev. Lett. {\bf 88}, 231803 (2002);
H. Albrecht {\it et al.} (ARGUS Collaboration), Phys. Lett. {\bf B 234}, 409 (1990);
B. Aubert {\it et al.} (BaBar Collaboration),   arXiv:hep-ex/0207081,
K. Abe {\it et al.} (Belle Collaboration), BELLE-CONF-0325.

\bibitem{bib:KEKB}
S.~Kurokawa and E.~Kikutani, 
Nucl. Instr. and Meth. {\bf A499}, 1 (2003),
and other papers included in this Volume.

\bibitem{bib:BELLE}
A.~Abashian {\it et al.} (Belle Collaboration), Nucl. Instr. and Meth. {\bf A479}, 117 (2002).

\bibitem{bib:GEANT}
R.~Brun, F.~Bruyant, M.~Maire, A.~C.~McPherson and P.~Zanarini,
CERN Report No. DD/EE/84-1 (1984).

\bibitem{bib:bu-model-lcsr} P.~Ball, arXiv:hep-ph/0306251.

\bibitem{bib:bu-model-isgw2} D.~Scora and N.~Isgur, Phys. Rev. {\bf D 52}, 
2783 (1995).

\bibitem{bib:bu-model-fn}
F.~De~Fazio and M.~Neubert, J.~High Energy Phys. {\bf 9906}, 017 (1999).

\bibitem{bib:bc-model} QQ $B$ meson event generator, developed by 
the CLEO Collaboration, http://www.lns.cornell.edu/public/CLEO/soft/QQ.

\bibitem{bib:limosani}
A. Limosani and T. Nozaki (Belle Collaboration), hep-ex/0407052.

\bibitem{bib:R2}
G.~C.~Fox and S.~Wolfram, Phys. Rev. Lett. {\bf 41}, 1581 (1978).

\bibitem{bib:cball} J.E.~Gaiser {\it et al.}, Phys. Rev. {\bf D 34}, 
711 (1986).

\bibitem{bib:argus} H.~Albrecht {\it et al.} (ARGUS Collaboration), 
Z.~Phys. {\bf C 48}, 543 (1990). 

\bibitem{bib:PDG2004} 
S. Eidelman {\it et al.}, Phys. Lett. {\bf B592 }, 1 (2004).

\bibitem{bib:patrick}
P.~Koppenburg {\it et al.} (Belle Collaboration), Phys. Rev. Lett. {\bf 93}, 061803  (2004).

\bibitem{bib:hfag-writeup} 
http:/www.slac.stanford.edu/xorg/hfag/semi/winter04/writeup.ps

\bibitem{bib:PDG2002}  
K.~Hagiwara {\it et al.}, Phys. Rev. {\bf D66 }, 010001 (2002).

\bibitem{bib:BaBar_vcb}
B.~Aubert {\it et al.}  (BaBar Collaboration),
Phys.\ Rev.\ Lett.\  {\bf 93}, 011803 (2004)

\bibitem{bib:inclusive}
R. Barate {\it et al.} (ALEPH Collaboration), Eur. Phys. J. {\bf C 6}, 555 (1999);
M. Acciarri {\it et al.} (L3 Collaboration), Phys. Lett. {\bf B 436} 174 (1998);
P. Abreu {\it et al.} (DELPHI Collaboration), Phys. Lett. {\bf B 478} 14 (2000);
G. Abbiendi {\it et al.} (OPAL Collaboration), Eur. Phys. J. {\bf C 21}, 399 (2001);
A. Bornheim {\it et al.} (CLEO Collaboration), CLEO-CONF 02-08.

\end{thebibliography}
\end{document}